\documentclass[10pt,lettersize,journal]{IEEEtran}
\usepackage{amsmath,amsfonts}
\usepackage{amssymb}
\usepackage{algorithmic}
\usepackage{algorithm}
\usepackage{array}
\usepackage{booktabs}
\usepackage{tabularx}
\usepackage[caption=false,font=normalsize,labelfont=sf,textfont=sf]{subfig}
\usepackage{textcomp}
\usepackage{stfloats}
\usepackage{url}
\usepackage{verbatim}
\usepackage{graphicx}
\usepackage{cite}
\begin{document}

\title{Riemannian Manifold Optimization for Advanced Wireless Communications: Fundamentals and Applications}

\author{ 
	Siwen Li,
	Jiacheng Chen,
    Yunting Xu,
    Shaofeng Li,
	Le Yao,
	Jieling Wang, and
	Dusit Niyato, \IEEEmembership{Fellow,~IEEE}
	
\thanks{
		\IEEEauthorblockA{S. Li and L. Yao are with the School of Cyber Science and Engineering, Southeast University, Nanjing 211102, China (email: siwenli@seu.edu.cn).} \\
		\indent\IEEEauthorblockA{J. Chen is with the Department of Strategic and Advanced Interdisciplinary Research, Pengcheng Laboratory, Shenzhen 518000, China (email: chenjch02@pcl.ac.cn).} \\
		\indent\IEEEauthorblockA{Y. Xu and D. Niyato are with the College of Computing and Data Science, Nanyang Technological University, Singapore, 639798 (e-mail: yunting.xu@ntu.edu.sg and dniyato@ntu.edu.sg).}\\
		\indent\IEEEauthorblockA{S. Li is with the School of Computer Science and Engineering, Southeast University, Nanjing 211102, China (email: shaofengli@seu.edu.cn).} \\
        \indent\IEEEauthorblockA{J. Wang is with 
		the State Key Laboratory of Integrated Services Networks, Xidian University, Xi'an 710068, China (email: jlwang@xidian.edu.cn).}\\
		\textit{(Corresponding author: Jiacheng Chen and Shaofeng Li.)}
	}
}



\maketitle

\begin{abstract}
Next-generation wireless communications promise transformative technologies such as massive multiple-input multiple-output (MIMO), reconfigurable intelligent surfaces (RIS), integrated sensing and communication (ISAC), and fluid antenna systems (FAS). 
However, deploying these technologies is hindered by large-scale optimization problems with nonconvex constraints. 
Conventional Euclidean-space methods rely on approximations or relaxations, which degrade performance and incur substantial computational costs. 
Riemannian manifold optimization (RMO) offers a powerful alternative that directly operates on the manifold defined by the geometric constraints. This approach inherently satisfies the constraints at every optimization step, thereby avoiding the performance degradation and substantial computational costs.
In this paper, we first elaborate on the principles of RMO, including the fundamental concepts, tools, and methods, emphasizing its effectiveness for nonconvex problems.
We then introduce its applications in advanced wireless communications, showing how constrained problems are reformulated on their natural manifolds and solved using tailored RMO algorithms. 
Furthermore, we present a case study on secure beamforming in an FAS-assisted non-orthogonal multiple access (NOMA) system, demonstrating RMO's superiority over conventional methods in terms of both performance and computational efficiency.
\end{abstract}


\section{Introduction}\label{sec:intro}
\IEEEPARstart{T}{he} sixth generation (6G) wireless communication system envisions extreme performance to enable applications such as holographic telepresence and immersive extended reality. The potential 6G technologies include large-scale massive multiple-input multiple-output (MIMO) \cite{10379539}, reconfigurable intelligent surfaces (RIS) \cite{10490002}, integrated sensing and communication (ISAC), movable antenna systems (MAS) and fluid antenna systems (FAS) \cite{10944486}. 
However, these technologies generally face formidable optimization challenges due to the inherent geometric constraints, such as the constant-modulus requirement on RIS phase shifters or the orthogonality constraint on precoding matrices. 
Conventional optimization techniques often struggle to address the complexity and geometric structure in these problems. 
For example, in Euclidean-space methods such as semidefinite relaxation (SDR), successive convex approximation (SCA), and majorization-minimization (MM) \cite{10636212}, relaxation can cause performance degradation, while iterative approximations impose substantial computational overheads. These limitations render them unsuitable for real-time, large-scale systems and underscore the need for a more direct, structure-preserving optimization framework.

To address these challenges, Riemannian manifold optimization (RMO) \cite{boumal2023introduction} has emerged as a powerful framework for handling the complex, nonlinear constraints inherent in next-generation wireless communication systems \cite{11040019, 8125771, 10542230, 10944479}. Rather than operating in a constrained Euclidean space, RMO optimizes directly on the manifold defined by the problem's geometric constraints. By exploiting the curved geometry of the feasible set—without relaxation or linearization—RMO inherently satisfies the constraints at every step. This preservation of structure often leads to higher computational efficiency and superior solutions, enabling RMO to outperform traditional methods in higher nonconvex landscapes.


RMO has already demonstrated notable success across diverse wireless communication tasks. In massive MIMO, it has been applied to beamforming and precoding designs to achieve higher throughput under strict power and interference limitations \cite{10886952}. In RIS systems, it efficiently determines phase configurations to maximize signal coverage \cite{10542230}; in ISAC, it effectively co-designs power allocation, beamforming, and radar waveforms \cite{10944479}. 
In our work \cite{11222098}, we have extended this geometric approach to the emerging field of FAS-assisted physical layer security, facilitating the joint optimization of antenna positions and secure beamforming vectors subject to transmit power constraints.
By reformulating these constraints on a product of complex spheres, RMO-based algorithms can solve the underlying subproblems without the performance loss and complexity associated with relaxation techniques.

\begin{figure*}[t]
	\centering
    \includegraphics[width=6.9 in]{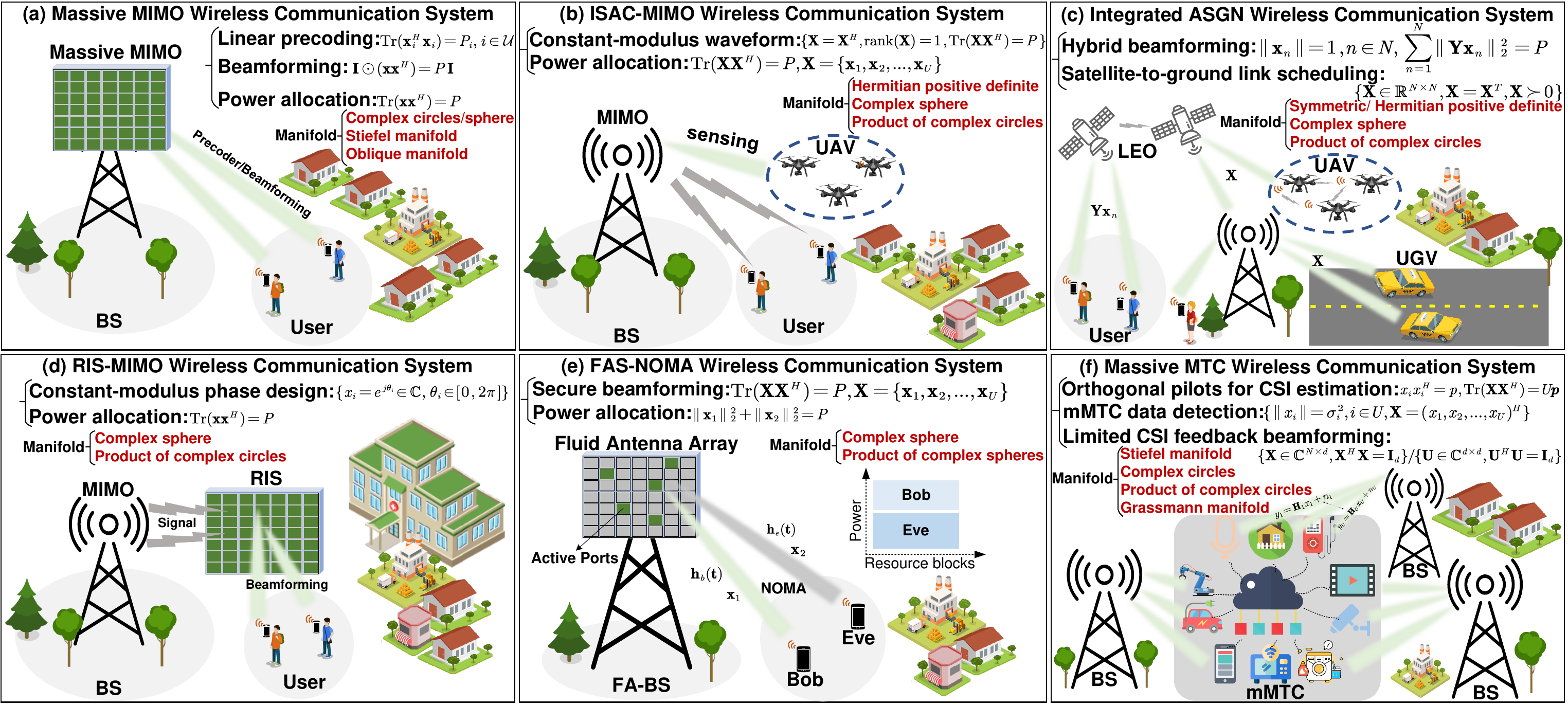} 
    \caption{Applications of RMO in advanced wireless communication systems. The figure illustrates several scenarios where RMO is applied to solve specific problems: (a) linear precoding, beamforming, and power allocation in massive MIMO systems; (b) constant modulus waveform design, phase design, and power allocation in ISAC-MIMO systems; (c) hybrid beamforming and link scheduling in ASGN; (d) phase shift design and power allocation in RIS-assisted massive MIMO systems; (e) secure beamforming and power allocation in FAS-assisted NOMA systems; and (f) signal detection, estimation, orthogonal pilot design, limited CSI feedback beamforming, and subspace signal design in mMTC systems.}
    \label{fig:RMO_in_Advanced_Wireless_Communication}
    \vspace{-5mm}
\end{figure*}

Motivated by these advancements, this article aims to present a comprehensive guide on RMO theory and its practical application in advanced wireless communications. 
As illustrated in Fig. \ref{fig:RMO_in_Advanced_Wireless_Communication}, RMO provides a versatile framework
for complex optimization problems in advanced wireless
communication systems by directly addressing nonconvex
constraints with geometric structure.
The contributions of this paper are as follows:
\begin{itemize}
    \item We introduce the fundamentals of RMO tailored for advanced wireless communications, covering its geometric foundations, essential optimization tools, and key algorithms, while identifying problem classes where it offers distinct advantages over conventional methods.
    \item We summarize applications of RMO to a range of prominent wireless communication problems, detailing their reformulation on suitable manifolds—such as the complex sphere, oblique, Stiefel, and Grassmann manifolds—and the design of tailored algorithms.
    \item We present a novel case study on FAS-assisted secure beamforming for non-orthogonal multiple access (NOMA) systems to demonstrate the practical advantages of RMO, showcasing its superior secrecy performance and computational efficiency.
    \item We discuss key considerations for implementing RMO in real-world systems and outline promising future research directions, highlighting its role as a pivotal enabler for next-generation wireless communication technologies.
\end{itemize}

The remainder of this paper is organized as follows. Section \ref{sec:riemannian} introduces the fundamentals of RMO, including its workflow, essential tools, and optimization methods.
Section \ref{sec:wireless_applications} summarizes the applications of RMO in advanced wireless communications, emphasizing how problems are mapped onto specific manifolds.
Section \ref{sec:case_study} presents a case study on secure beamforming in FAS-assisted NOMA systems, demonstrating the advantages of RMO.
Section \ref{sec:future} discusses future research directions for RMO in wireless communications.
Section \ref{sec:conclusion} concludes the paper.

\section{Fundamentals of RMO}\label{sec:riemannian}
\subsection{The Workflow of RMO}\label{sec:euclidean_to_riemannian}
A manifold is a topological space that locally resembles Euclidean space. This local Euclidean property allows for the application of calculus, making it possible to define concepts like smoothness, differentiation, and integration. An embedded submanifold is a manifold contained within a higher-dimensional Euclidean space, where its geometry is inherited from the ambient space. A Riemannian manifold is a real smooth manifold equipped with a smoothly varying inner product on the tangent space at each point. This structure allows for the definition of geometric notions such as angles, lengths of curves, and gradients \cite{boumal2023introduction}.

In wireless communications, optimization problems often involve constraints that naturally define a manifold structure. This geometric structure arises directly from the problem's physical or mathematical constraints, such as the constant amplitude of signals, orthogonality between precoding vectors, or the positive definite nature of covariance matrices. This geometric structure is fundamental to a wide range of modern applications, as depicted in Fig.~\ref{fig:RMO_in_Advanced_Wireless_Communication}, including precoding and beamforming in massive MIMO; waveform design in ISAC; hybrid beamforming and link scheduling in air-space-ground networks (ASGN); constant-modulus phase design in RIS; secure communications in FAS-assisted NOMA; and orthogonal pilot design in massive machine-type communications (mMTC). Identifying the appropriate manifold is a crucial first step in all such problems. The common structures underlying these applications include: the complex circle manifold for constant-modulus constraints; the sphere and oblique manifolds for beamforming; the Stiefel manifold of orthonormal matrices for orthogonal precoding; the Grassmann manifold of subspaces for limited channel state information (CSI) feedback systems; the manifold of positive definite matrices for covariance optimization; and product manifolds for problems with multiple constraints.

\begin{figure*}[t]
    \centering
    \includegraphics[width=7.0 in]{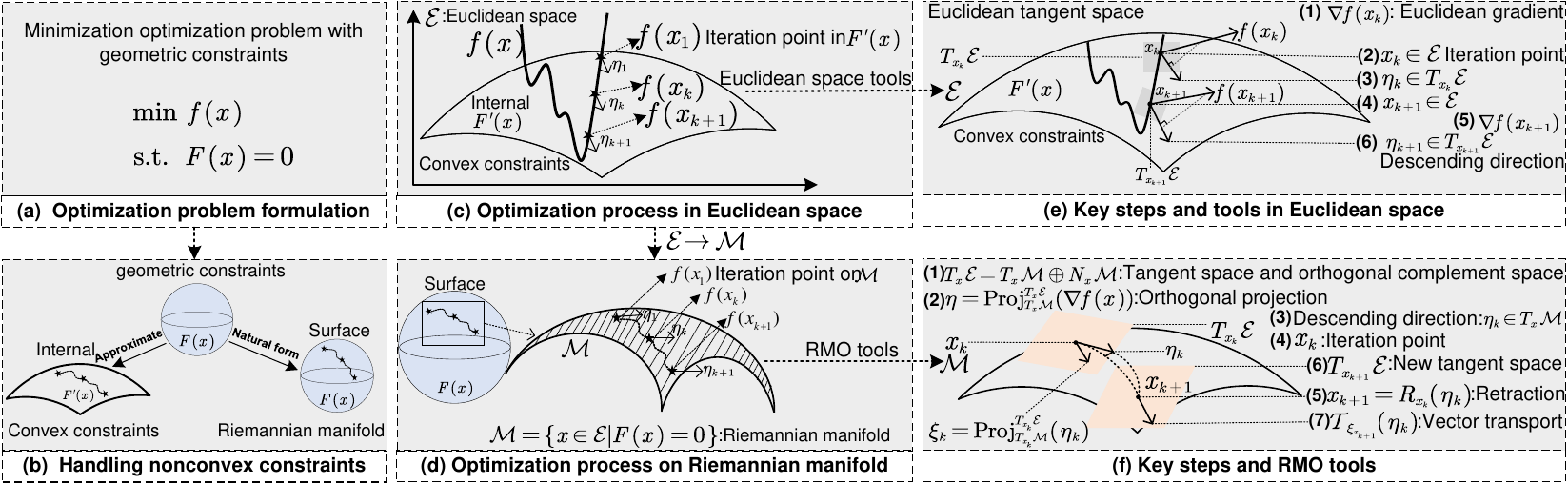} 
    \caption{Illustration of a Riemannian optimization step. The figure depicts the core components of an iterative update in Euclidean space and on a manifold. Starting from the current iteration, the Euclidean gradient is first projected onto the tangent space to obtain the Riemannian gradient, which defines a search direction. A step is then taken within the tangent space, followed by a retraction that maps the resulting point back onto the manifold to produce the next iteration, ensuring that feasibility is maintained.}
    \label{fig:Euclidean_to_RM}
    \vspace{-5mm}
\end{figure*}

RMO leverages the inherent geometry by reformulating the problem as an unconstrained optimization on the manifold itself.
As illustrated in Fig.~\ref{fig:Euclidean_to_RM}, a typical RMO algorithm is an iterative process that leverages the geometry of the manifold by operating within its tangent spaces. For embedded submanifolds, this process elegantly interfaces with the ambient Euclidean space and involves four key steps: (i) computing the standard Euclidean gradient of the cost function; (ii) orthogonally projecting the Euclidean gradient onto the tangent space at the current iteration to obtain the Riemannian gradient; (iii) taking a step along this new search direction within the tangent space; and (iv) applying a retraction to map the resulting point from the tangent space back onto the manifold, which produces the next feasible iteration. This process is repeated until the norm of the Riemannian gradient falls below a predefined tolerance. More advanced algorithms, such as the conjugate gradient method, employ vector transport to move tangent vectors between tangent spaces, while second-order methods use the Riemannian Hessian to utilize curvature information. 
RMO avoids the need to project infeasible iterations back onto the manifold, a potentially costly step in conventional methods. Instead, it uses computationally efficient retractions to guarantee feasibility at each step, often leading to more robust convergence \cite{10636212,boumal2023introduction}.

\subsection{RMO Tools}\label{sec:riemannian_optimization_tools}
In Euclidean space, optimization proceeds along straight lines. However, when solutions are confined to a curved manifold, such as a sphere, standard gradient steps may yield infeasible points. To address this, RMO operates directly on the manifold \cite{boumal2023introduction}.
The iterative process illustrated in Fig.~\ref{fig:Euclidean_to_RM}—projecting a gradient, moving within a tangent space, and retracting back to the manifold—cannot be executed with standard Euclidean operations. Performing these steps requires a specialized geometric toolkit. The key tools are defined below:
\begin{enumerate}
    \item \textbf{Tangent Space:} At any point on the manifold, the tangent space is a real vector space that serves as a first-order linear approximation of the manifold. It contains all feasible search directions, and local computations, such as determining a descent direction, are performed within this space.
    \item \textbf{Riemannian Metric:} A Riemannian metric defines a smoothly varying inner product on each tangent space. This metric endows the manifold with a geometric structure, allowing for the measurement of distances, and is essential for defining the Riemannian gradient.
    \item \textbf{Riemannian Gradient:} For a smooth cost function, the Riemannian gradient is the tangent vector indicating the steepest descent direction under the Riemannian metric. For minimization, moving in the direction of the negative gradient yields the steepest descent. For embedded manifolds, it is obtained by projecting the Euclidean gradient onto the tangent space.
    \item \textbf{Retraction:} A retraction is a computationally efficient mapping that translates a step in a tangent space into a new feasible iteration on the manifold. Instead of calculating the exact geodesic (the shortest path along the manifold's surface), which is often costly, a retraction offers a simpler, first-order approximation. This approach is sufficient to guarantee convergence for many algorithms, striking a practical balance between geometric accuracy and computational efficiency.
    \item \textbf{Vector Transport:} Since tangent spaces at different points are distinct, vectors cannot be directly combined. Vector transport addresses this by moving a vector from one tangent space to another, an operation crucial for algorithms that accumulate information across iterations, such as quasi-Newton methods.
\end{enumerate}

\subsection{RMO Methods}\label{sec:riemannian_optimization_methods}
By leveraging the RMO tools defined above, classical optimization algorithms can be generalized to operate on manifolds. These tools form the building blocks for iterative methods that efficiently solve constrained problems by ensuring all iterations remain on the manifold. The main algorithm classes are distinguished by the information they use to determine the search direction, which creates a trade-off between per-iteration cost and convergence speed:
\begin{enumerate}
    \item \textbf{First-Order Methods:} These methods rely solely on gradient information. The most fundamental is Riemannian gradient descent (RGD), which uses the Riemannian gradient to define a descent direction at each iteration. While computationally efficient, RGD often exhibits slow, linear convergence. A more advanced first-order method is the Riemannian conjugate gradient (RCG), which improves upon RGD by incorporating information from the previous search direction. By using vector transport to create a sequence of conjugate search directions, RCG can achieve a much faster, often super-linear, convergence rate without needing to compute or store Hessian information.
    
    \item \textbf{Second-Order Methods:} To achieve faster convergence, methods like the Riemannian trust-region (RTR) algorithm incorporate second-order (curvature) information by using the Riemannian Hessian. They build a more accurate quadratic model of the cost function within the tangent space to find a more effective step. This use of the Hessian enables a much faster quadratic convergence rate but at a high computational cost, making them less practical for large-scale problems.
    
    \item \textbf{Quasi-Newton Methods:} Offering a balance between first- and second-order approaches, quasi-Newton methods achieve super-linear convergence without the full cost of the Hessian. These methods rely on vector transport to accumulate past gradient information, which is used to build an efficient approximation of the Hessian, achieving super-linear convergence without the full computational cost of the Hessian. This approach makes these methods a popular choice for many practical applications \cite{boumal2023introduction}.
\end{enumerate}

With this foundational understanding of RMO's geometric tools and algorithmic framework, we are now prepared to map these concepts onto the specific manifold structures that arise in key wireless communication problems.

\begin{figure*}[ht]
    \centering
    \includegraphics[width=7.0 in]{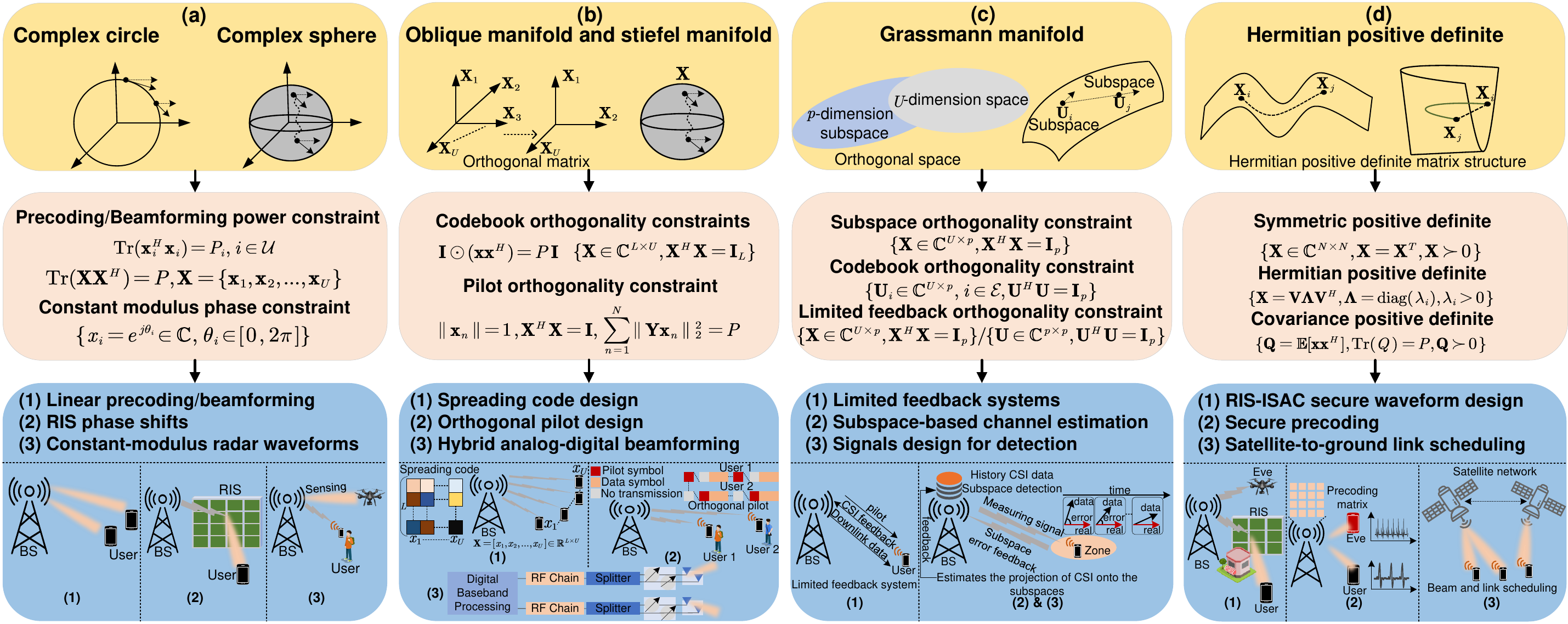} 
    \caption{An illustration of typical manifold structures in wireless communication. The figure showcases four representative classes of manifolds, each defined by specific geometric constraints, and maps them to key applications: (a) complex circle/sphere manifolds for precoding, beamforming, and constant-modulus waveform design; (b) the oblique and Stiefel manifolds for designing spreading codes and orthogonal pilots; (c) the Grassmann manifold for limited feedback, subspace-based estimation, and noncoherent signal design; and (d) the Hermitian positive definite manifold for secure precoding, RIS-assisted ISAC waveform design, and satellite link scheduling.}
    \label{fig:RMO_in_Wireless_Communication_Applications}
    \vspace{-5mm}
\end{figure*}

\section{Applications in Wireless Communications}\label{sec:wireless_applications} 
This section details how different wireless communication problems can be mapped onto specific manifold structures, as well as the design of geometrically consistent and efficient algorithms.

\subsection{Mapping Wireless Problems onto Manifold Structures}
Fig.~\ref{fig:RMO_in_Wireless_Communication_Applications} visualizes the geometric structures for several key wireless communication applications and shows how the RMO framework maps optimization problems onto their corresponding manifolds. 
The following subsections are organized according to the four representative manifold classes depicted in the figure, detailing their geometric formulations and corresponding applications.

\subsubsection{Constant-Modulus and Constant-Norm Optimization}
Many wireless communication systems impose constant-modulus or constant-norm constraints, often due to hardware limitations such as power amplifiers operating near saturation. These problems can be modeled on the product of complex circles or the complex sphere manifold, as visualized in Fig.~\ref{fig:RMO_in_Wireless_Communication_Applications}(a). RMO provides an effective framework for these scenarios, enabling the direct optimization of energy-efficient precoding and beamforming vectors \cite{8125771}, phase shifts for RIS to enhance communication quality \cite{10542230}, and radar waveforms with a low peak-to-average-power ratio for improved detection in ISAC systems \cite{10944479}. By operating directly on the manifold, RMO intrinsically satisfies the constraints, avoiding the performance loss associated with projection or relaxation methods.

\subsubsection{Unit-Norm and Orthonormal Column Constraints}
Constraints involving matrices with unit-norm or orthonormal columns are common in tasks requiring resource separation and interference mitigation. These problems are naturally formulated on the oblique manifold (for unit-norm columns) and the Stiefel manifold (for orthonormal columns), as depicted in Fig.~\ref{fig:RMO_in_Wireless_Communication_Applications}(b). The key applications include hybrid analog-digital beamforming, where the analog precoder is constrained to be unitary to ensure beam orthogonality \cite{10886952}, and the design of orthogonal pilot sequences or spreading codes to minimize interference in channel estimation and multiuser access. Traditional methods often require computationally expensive and numerically sensitive re-orthogonalization steps. In contrast, RMO algorithms inherently preserve orthogonality through their update steps (i.e., retractions), leading to more efficient and numerically stable optimization.

\subsubsection{Subspace-Based Optimization} 
In certain applications, the optimization variable is a subspace rather than a specific basis, making the problem invariant to basis representation. As illustrated in Fig.~\ref{fig:RMO_in_Wireless_Communication_Applications}(c), the Grassmann manifold represents the set of all subspaces of a given dimension and provides the ideal geometric setting for such problems. This is particularly useful in limited feedback systems for quantizing beamforming matrices or CSI into codebooks, as the manifold structure allows for the design of efficient, basis-independent distance metrics \cite{10649656}. Further applications include subspace-based channel estimation for improved accuracy and the design of signals for noncoherent detection in covert communications over fading channels \cite{10879056}. RMO on the Grassmann manifold eliminates basis redundancy, which improves the numerical conditioning of the problem.

\subsubsection{Covariance Matrix Optimization} Optimization problems involving covariance matrices, which must be Hermitian positive definite (HPD), are prevalent in statistical signal processing, as shown in Fig.~\ref{fig:RMO_in_Wireless_Communication_Applications}(d). The manifold of HPD matrices provides the natural space for these problems, ensuring this essential structure is preserved throughout the optimization process \cite{11018229}. This approach is critical for ISAC waveform design, where the goal is to jointly optimize communication and sensing performance by shaping the transmit covariance matrix, as well as for secure precoding, which involves designing covariance matrices for signals and artificial noise \cite{11040019}. Its real-valued counterpart, the manifold of symmetric positive definite (SPD) matrices, has been used to model spatio-temporal traffic correlations for machine learning-based link scheduling in satellite networks \cite{10851312}. In both the complex and real cases, RMO avoids the risk of iterations losing their positive definite structure—a common issue in conventional methods that often require costly projections or relaxations. By operating directly on the manifold, RMO inherently guarantees feasibility and can lead to improved convergence.

\subsection{Algorithm Design on Manifolds}
Once a problem is formulated on a manifold, the choice of algorithm is a critical design step, guided by the trade-off between per-iteration cost and convergence speed. 
For applications prioritizing low latency, such as real-time RIS optimization on the complex circle manifold, first-order methods are a natural fit. For instance, the work in \cite{10542230} employs a parallel RCG algorithm to jointly optimize the RIS phase shifts and transmit power. 
This approach achieves an approximately quadratic computational complexity with respect to the number of reflecting elements, a significant improvement over the third-order complexity of the conventional MM benchmark.
While the simpler RGD method offers linear complexity, its convergence can be slow. In contrast, RCG strikes a superior balance by leveraging previous search directions to accelerate convergence without Hessian computations. In symbol detection, for instance, it attains a complexity that is approximately linear in the number of antennas and users, outperforming the third-order complexity of Euclidean methods \cite{8125771}. Similarly, in cell-free ISAC, an RCG-based beamforming design achieves a complexity that scales linearly with system dimensions, significantly improving upon the third-order complexity of SDR \cite{10944479}.

Additionally, for problems where faster convergence is paramount and a higher per-iteration cost is acceptable—such as orthogonal beamforming on the Stiefel manifold and covariance matrix optimization—quasi-Newton and second-order methods are more effective.
Quasi-Newton methods achieve super-linear convergence by building an efficient approximation of the Hessian. For instance, our previous work \cite{11222098} developed a quasi-Newton method to optimize secure beamforming in a FAS-assisted multi-user system. This approach exhibits an approximately quadratic computational complexity with antenna numbers, offering a substantial improvement over the third-order complexity of the SDR-based benchmark. When high precision is paramount or the problem geometry is particularly challenging, second-order methods like the RTR algorithm become indispensable \cite{11018229}. 
For offline tasks, such as covariance matrix optimization on the HPD manifold \cite{11040019, 10851312}, the trade-off of using second-order methods can be advantageous. Although computing the Riemannian Hessian is costly per iteration, the quadratic convergence rate can drastically reduce the required iterations. This strategic tailoring of algorithms to specific problem structures is a key strength of the RMO framework.

\section{Case Study}\label{sec:case_study}
This section examines the problem of secure beamforming in an FAS-assisted NOMA system. This case study focuses on an FAS-assisted system to underscore a critical challenge in next-generation wireless communications: leveraging new physical degrees of freedom requires highly efficient optimization algorithms. The core advantage of FAS is its ability to continuously reconfigure the wireless channel by adjusting antenna positions. However, finding optimal positions requires evaluating secure beamforming performance for numerous candidates. Conventional optimization methods for beamforming are often inefficient, rendering FAS deployment impractical. Thus, this case study demonstrates that RMO's computational efficiency is essential for enabling joint optimization in FAS.

\subsection{System Model and Problem Formulation}
We consider a downlink scenario where a base station (BS) equipped with a fluid antenna (FA) array transmits a confidential signal to a single-antenna legitimate user (Bob) in the presence of a potential eavesdropper (Eve). The FA array can dynamically reposition within a compact two-dimensional region, thereby flexibly manipulating channel conditions, as shown in Fig. \ref{fig:RMO_in_Advanced_Wireless_Communication}(e).
Unlike conventional fixed-position antenna systems, the channel from the BS to Bob depends on the antenna position. The objective is to maximize the secrecy rate—defined as the difference between the achievable rates of Bob and Eve—by jointly optimizing the beamforming vector and the antenna positions, subject to the BS's total transmit-power constraint and the allowable movement region.

\begin{figure}[t]
\centering
\includegraphics[width=3.0 in]{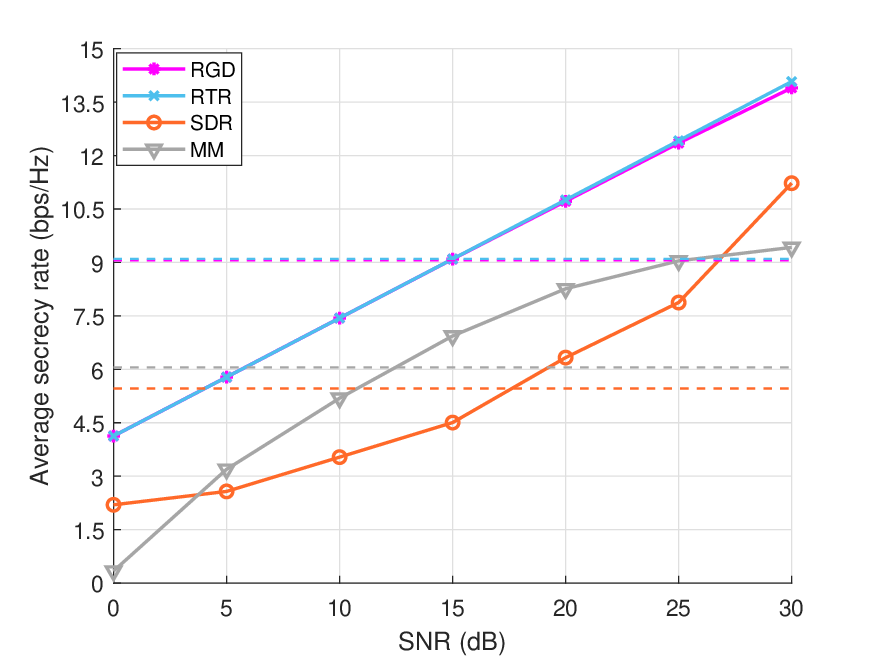} 
\caption{Average secrecy rate versus SNR.}
\label{fig:ASR_SNR}
\vspace{-4mm}
\end{figure}
\begin{figure}[t]
    \centering
    \includegraphics[width=3.0 in]{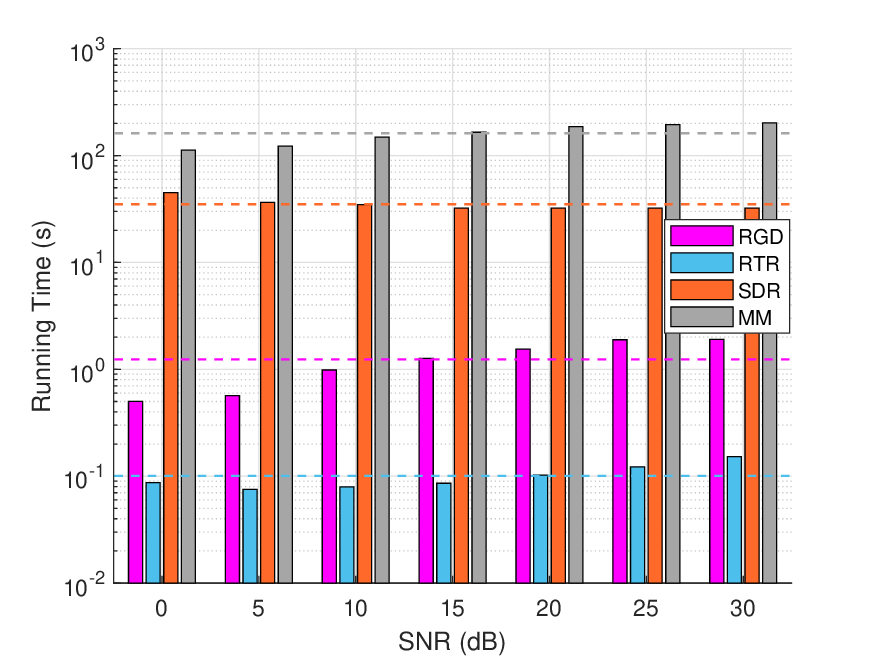} 
    \caption{Runtime to convergence versus SNR.}
    \label{fig:Running_Time}
    \vspace{-3mm}
\end{figure}

\subsection{RMO Solution and Baseline Methods}
For any given set of FA positions, the secure beamforming problem must be solved to evaluate the system's performance. The transmit power constraints on the beamforming vectors confine them to a complex spheres manifold, which is a Riemannian manifold. Instead of adopting conventional SDR or MM approaches \cite{10636212} that relax these constraints, we formulate the beamforming problem directly on this manifold. 
This problem is then addressed using two representative RMO algorithms: the first-order RGD and the second-order RTR. 
Both methods treat the secrecy rate as the cost function on the manifold, where the RGD algorithm iteratively updates the beamforming vectors along the negative Riemannian gradient.
In contrast, the more advanced RTR method leverages both the Riemannian gradient and Hessian to form a local quadratic model of the problem, typically achieving faster convergence. Crucially, both methods employ a retraction after each update to ensure the iterates remain on the manifold, thereby strictly satisfying the power constraints. 
The efficiency of this RMO formulation is critical, as it directly determines the feasibility of the search for optimal FA positions. By preserving the problem's geometric structure, RMO avoids relaxation-induced errors, ensures every iteration is feasible, and achieves rapid convergence.

To evaluate their performance, the proposed RMO methods are benchmarked against two conventional baseline schemes. The first is the SDR method, which transforms the problem by relaxing the nonconvex power constraints into a convex semidefinite program solvable via standard interior-point methods. The second is the MM algorithm, which iteratively tackles a sequence of more tractable surrogate problems, each of which is convex and solved by CVX.

\subsection{Performance Evaluation and Insights}
Fig. \ref{fig:ASR_SNR} illustrates the average secrecy rate (ASR) versus the signal-to-noise ratio (SNR) for a system where the BS is equipped with 8 FA ports and 4 FAs. 
The dotted line represents the ASR value of each algorithm averaged over the entire SNR range.
The proposed RTR and RGD algorithms demonstrate superior ASR performance, significantly outperforming the baseline schemes. 
Averaged over the entire SNR range, the proposed RTR algorithm demonstrates a significant performance advantage. It achieves ASR improvement of approximately $66.57\%$ and $50.27\%$ compared to the MM and SDR methods, respectively. 
While RTR achieves the best results, RGD also yields strong performance, confirming the overall effectiveness of the RMO approach for secure beamforming design.
This substantial improvement highlights the benefits of manifold optimization, which avoids the performance degradation caused by approximation and relaxation errors in the MM and SDR approaches, respectively. 
As the SNR increases, the performance gap between RTR and RGD widens slightly, as RTR utilizes second-order information while RGD relies only on first-order information.
Additionally, the SDR method underperforms the MM method in the low-SNR region but surpasses it at high SNRs, because the projection of the SDR solution onto the feasible set is more accurate than the MM approximation, especially in the high-SNR regime. 
Notably, the MM method's performance saturates due to inherent approximation errors becoming more pronounced at higher SNRs, leading to suboptimal secure beamforming solutions.
These results demonstrate the robustness and superior performance of RMO-based methods compared to Euclidean space-based approaches.

Fig. \ref{fig:Running_Time} compares the runtime to convergence for the different algorithms \footnote{This runtime is obtained by running the code on a Windows computer with a 3.20 GHz i9-14900K CPU and 32 GB RAM.}. The dotted line represents the average runtime to convergence value of each algorithm over the entire SNR range.
In terms of average runtime, the RTR algorithm is substantially more efficient, requiring only 0.10 seconds to converge. In contrast, the RGD, SDR, and MM algorithms take significantly longer, with average convergence times of 1.23, 35.01, and 161.93 seconds, respectively. 
While each RTR step is costlier, its quadratic convergence significantly reduces the total iteration count, making it substantially faster than RGD overall.
The MM algorithm's runtime exceeds that of the SDR algorithm, indicating that its iterative approximation process is more computationally demanding. Furthermore, both MM and SDR exhibit high runtimes, making them unsuitable for large-scale, real-time systems. In contrast, the efficiency of RMO-based solvers enables exploration of the FAS design space, facilitating the performance gains of this technique.


\vspace{-2mm}
\section{Future Directions}\label{sec:future}
As wireless communication systems continue to evolve, the role of RMO is poised to expand, driven by its unique ability to handle increasingly complex geometric constraints. This section discusses future directions for RMO in wireless communications.

\vspace{-2mm}
\subsection{RMO for AI-Enabled Wireless Systems}
In artificial intelligence (AI)-enabled wireless systems, data-driven solutions like neural network-generated precoders must satisfy physical hardware constraints.
For instance, neural networks for RIS optimization must generate phase shifts on the product of complex circles, while those for hybrid beamforming must produce precoders on the Stiefel manifold. Conventional training with Euclidean optimizers often inefficiently enforces these constraints, requiring repeated projections that can hinder convergence and yield suboptimal results. 
RMO offers a natural solution by reformulating the training process as an unconstrained optimization problem directly on the manifold.
This geometry-aware training, using optimizers like Riemannian stochastic gradient descent, inherently satisfies the physical constraints at each iteration. 
Furthermore, manifold learning can enhance this framework by reducing channel data dimensionality while preserving its geometric structure, thus accelerating convergence and improving performance.

\vspace{-2mm}
\subsection{Large-Scale Low Earth Orbit Satellite Communications}
Large-scale low-earth-orbit (LEO) satellite constellations face complex optimization challenges, such as real-time resource management, arising from high mobility, dynamic topologies, and strict latency constraints. RMO offers a robust framework for these problems, as key variables like satellite antenna orientations (sphere manifold) and beamforming vectors (Stiefel manifold) are inherently geometric. This enables the development of efficient, real-time algorithms that operate directly on these manifolds, enhancing spectral efficiency and network resilience.


\vspace{-2mm}
\subsection{Joint Hardware Design and Signal Processing}
A key future direction is the co-design of physical hardware parameters and signal processing algorithms. This paradigm jointly optimizes variables across both domains, for instance, by treating RIS phase shifts or holographic MIMO surface currents as optimizable variables. Such problems introduce complex, often discrete or mixed continuous-discrete, search spaces that are coupled with conventional manifold constraints.
In this cross-layer optimization, RMO efficiently solves the manifold-constrained signal processing subproblems for each candidate hardware configuration.
Extending RMO to non-smooth or discrete manifolds is a promising research avenue. Such an extension could unify these coupled problems into a single geometry-aware framework, enabling a holistic system design with significant gains in performance and efficiency.

\vspace{-2mm}
\section{Conclusion}\label{sec:conclusion}
In this paper, we have provided a comprehensive guide to the fundamentals and applications of RMO in advanced wireless communication systems. We have shown that by leveraging the inherent geometric structure of constraints, RMO offers a versatile and efficient framework that overcomes limitations of conventional Euclidean-space methods so as to tackle the complex optimization challenges in modern and future wireless communications. 
We have detailed how key wireless communication problems can be reformulated on specific manifolds and presented a case study on FAS-assisted secure beamforming validating the superior performance and computational efficiency of the RMO approach.
Finally, we have outlined promising future research directions, including the integration of RMO with AI-driven systems, its application to LEO satellite communications, and its potential role in joint hardware-signal processing co-design. 

\bibliographystyle{IEEEtran}
\bibliography{refs.bib}

\begin{thebibliography}{10}
\providecommand{\url}[1]{#1}
\csname url@samestyle\endcsname
\providecommand{\newblock}{\relax}
\providecommand{\bibinfo}[2]{#2}
\providecommand{\BIBentrySTDinterwordspacing}{\spaceskip=0pt\relax}
\providecommand{\BIBentryALTinterwordstretchfactor}{4}
\providecommand{\BIBentryALTinterwordspacing}{\spaceskip=\fontdimen2\font plus
\BIBentryALTinterwordstretchfactor\fontdimen3\font minus \fontdimen4\font\relax}
\providecommand{\BIBforeignlanguage}[2]{{%
\expandafter\ifx\csname l@#1\endcsname\relax
\typeout{** WARNING: IEEEtran.bst: No hyphenation pattern has been}%
\typeout{** loaded for the language `#1'. Using the pattern for}%
\typeout{** the default language instead.}%
\else
\language=\csname l@#1\endcsname
\fi
#2}}
\providecommand{\BIBdecl}{\relax}
\BIBdecl

\bibitem{10379539}
Z.~Wang, J.~Zhang, H.~Du, D.~Niyato, S.~Cui, B.~Ai, M.~Debbah, K.~B. Letaief, and H.~V. Poor, ``A tutorial on extremely large-scale mimo for 6g: Fundamentals, signal processing, and applications,'' \emph{IEEE Communications Surveys \& Tutorials}, vol.~26, no.~3, pp. 1560--1605, 2024.

\bibitem{10490002}
X.~Zhao, H.~Liu, S.~Gong, X.~Ju, C.~Xing, and N.~Zhao, ``Dual-functional mimo beamforming optimization for ris-aided integrated sensing and communication,'' \emph{IEEE Transactions on Communications}, vol.~72, no.~9, pp. 5411--5427, 2024.

\bibitem{10944486}
C.~Jiang, C.~Zhang, C.~Huang, J.~Ge, D.~Niyato, and C.~Yuen, ``Movable antenna-assisted integrated sensing and communication systems,'' \emph{IEEE Transactions on Wireless Communications}, vol.~24, no.~8, pp. 6397--6412, 2025.

\bibitem{10636212}
Y.-F. Liu, T.-H. Chang, M.~Hong, Z.~Wu, A.~Man-Cho~So, E.~A. Jorswieck, and W.~Yu, ``A survey of recent advances in optimization methods for wireless communications,'' \emph{IEEE Journal on Selected Areas in Communications}, vol.~42, no.~11, pp. 2992--3031, 2024.

\bibitem{boumal2023introduction}
N.~Boumal, \emph{An introduction to optimization on smooth manifolds}.\hskip 1em plus 0.5em minus 0.4em\relax Cambridge University Press, 2023.

\bibitem{11040019}
A.~Pourkabirian, W.~Ni, X.~Zhou, K.~Li, and M.~H. Anisi, ``A precoding perturbation method in geometric optimization: Exploring manifold structure for privacy and efficiency,'' \emph{IEEE Transactions on Information Forensics and Security}, vol.~20, pp. 6220--6235, 2025.

\bibitem{8125771}
J.-C. Chen, ``Manifold optimization approach for data detection in massive multiuser mimo systems,'' \emph{IEEE Transactions on Vehicular Technology}, vol.~67, no.~4, pp. 3652--3657, 2018.

\bibitem{10542230}
Y.~Geng, T.~Hiang~Cheng, K.~Zhong, and K.~Chan~Teh, ``Unified manifold optimization for double-irs-aided mimo communication,'' \emph{IEEE Communications Letters}, vol.~28, no.~7, pp. 1713--1717, 2024.

\bibitem{10944479}
S.~Zargari, D.~Galappaththige, C.~Tellambura, and G.~Ye~Li, ``Downlink beamforming for cell-free isac: A fast complex oblique manifold approach,'' \emph{IEEE Transactions on Wireless Communications}, vol.~24, no.~8, pp. 6458--6474, 2025.

\bibitem{10886952}
R.~Sun, L.~You, A.-A. Lu, C.~Sun, X.~Gao, and X.-G. Xia, ``Precoder design for user-centric network massive mimo with matrix manifold optimization,'' \emph{IEEE Journal on Selected Areas in Communications}, vol.~43, no.~3, pp. 705--719, 2025.

\bibitem{11222098}
S.~Li, J.~Chen, S.~Ren, B.~Deng, J.~Wang, and J.~Wang, ``Low-complexity secure beamforming with fluid antenna-assisted mu-miso system,'' \emph{IEEE Transactions on Information Forensics and Security}, pp. 1--1, 2025.

\bibitem{10649656}
Y.~Cao, H.~Yin, Z.~Qin, W.~Li, W.~Wu, and M.~Debbah, ``A manifold learning-based csi feedback framework for fdd massive mimo,'' \emph{IEEE Transactions on Communications}, vol.~73, no.~3, pp. 1833--1846, 2025.

\bibitem{10879056}
H.~Fukada, H.~Iimori, C.~Pradhan, S.~Malomsoky, and N.~Ishikawa, ``Covert communications without pre-sharing of side information and channel estimation over quasi-static fading channels,'' \emph{IEEE Wireless Communications Letters}, vol.~14, no.~4, pp. 1239--1243, 2025.

\bibitem{11018229}
B.~Li, Y.~Hu, Z.~Dong, E.~Panayirci, H.~Jiang, and Q.~Wu, ``Energy-efficient secure design for ios and an aided cf-mmimo network,'' \emph{IEEE Transactions on Wireless Communications}, pp. 1--1, 2025.

\bibitem{10851312}
J.~J. Sadique, I.~Nasim, and A.~S. Ibrahim, ``Link scheduling in satellite networks via machine learning over riemannian manifolds,'' \emph{IEEE Open Journal of the Communications Society}, vol.~6, pp. 972--985, 2025.

\end{thebibliography}

\end{document}